\documentclass[aps,pra,showpacs,superscriptaddress,twocolumn]{revtex4}
\usepackage{graphicx,epsfig,color}
\usepackage{amssymb}
\usepackage{amsmath}
\usepackage{bm}

\begin{document}
\title{Optical pumping effect in absorption imaging of $F=1$ atomic gases}
\author{Sooshin Kim}
\author{Sang Won Seo}
\affiliation{Department of Physics and Astronomy, and Institute of Applied Physics, Seoul National University, Seoul 08826, Korea}
\author{Heung-Ryoul Noh}\email{hrnoh@chonnam.ac.kr}
\affiliation{Department of Physics, Chonnam National University, Gwangju 61186, Korea}
\author{Y. Shin}\email{yishin@snu.ac.kr}
\affiliation{Department of Physics and Astronomy, and Institute of Applied Physics, Seoul National University, Seoul 08826, Korea}
\affiliation{Center for Correlated Electron Systems, Institute for Basic Science, Seoul 08826, Korea}


\begin{abstract}
We report our study of the optical pumping effect in absorption imaging of $^{23}$Na atoms in the $F=1$ hyperfine spin states. Solving a set of rate equations for the spin populations in the presence of a probe beam, we obtain an analytic expression for the optical signal of the $F=1$ absorption imaging. Furthermore, we verify the result by measuring the absorption spectra of $^{23}$Na Bose-Einstein condensates prepared in various spin states with different probe beam pulse durations. The analytic result can be used in the quantitative analysis of $F=1$ spinor condensate imaging and readily applied to other alkali atoms with $I=3/2$ nuclear spin such as $^{87}$Rb.
\end{abstract}

\maketitle

\section{Introduction}\label{section1}

Optical imaging is the only successful method for scrutinizing atomic samples in ultracold atom experiments and various techniques such as absorption imaging~\cite{Anderson,Andrews1}, phase-contrast imaging~\cite{Andrews,Higbie}, and fluorescence imaging~\cite{Sherson10,Parsons15} have been developed for different purposes. One presumed requirement of imaging is that the sample condition in terms of atomic density and spin state should be unchanged during imaging. If it is not the case, the optical signal needs to be interpreted in the imaging process so as to take into account the temporal evolution of the sample condition, which would be a somewhat complicated task. To fulfill the requirement in experiments, a short-pulse probe beam or a cyclic optical transition may be used for imaging. However, a short-pulse probe beam might suffer from a low signal-to-noise ratio, and optical pre-pumping for the cyclic transition would perturb the sample by imparting photon recoil momentum and washing out spin-state information.

In this paper, we investigate absorption imaging of alkali atoms with $I=3/2$ nuclear spin in the $F=1$ hyperfine ground state. This is a common situation in ultracold atom experiments with bosonic atoms such as $^{23}$Na and $^{87}$Rb. We focus particularly on the optical pumping effect by non-cyclic transitions in $F=1$ imaging, i.e., atoms in the $F=1$ spin states are pumped into the $F=2$ spin states, which are optically dark for the probe beam because of large frequency detuning~\cite{Metcalf}. Figure 1 shows examples for absorption images of $F=1$ Bose-Einstein condensates (BECs), which were taken with different pulse durations of the probe beam. Because of the population depletion of the $F=1$ spin states induced by the probe beam, the measured optical depth decreases as the pulse duration increases~\cite{Pappas,Han}. A quantitative interpretation of the imaging signal requires precise knowledge of the temporal evolution of the atomic spin state during imaging. Here, we solve a set of rate equations for the spin populations in the presence of a probe beam and obtain an analytic expression for the optical signal of $F=1$ absorption imaging. We verify the analytic result by experimentally measuring the absorption spectra of $^{23}$Na BECs prepared in various spin states with different probe-beam pulse durations. The analytic solution of spin population evolution under imaging light can be readily used in the quantitative analysis of other imaging methods for $F=1$ spinor BECs~\cite{Higbie,Seo}.

\begin{figure}
\centering
\includegraphics[width=8.2cm]{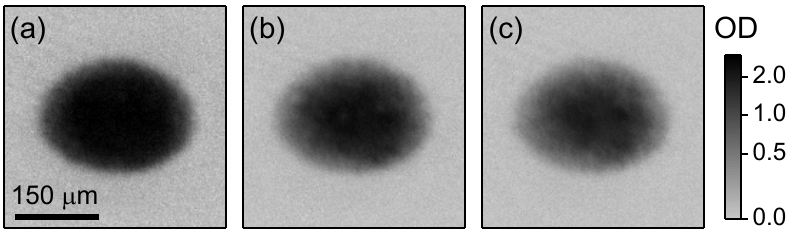}
\caption{Absorption images of Bose-Einstein condensates of $^{23}$Na atoms in the $|F_g=1, m_F=-1\rangle$ state. The probe beam was resonant on the $|F_g=1\rangle \rightarrow |F_e=2\rangle$ $D_2$ transition and its pulse duration was (a) $\tau_0=20~\mu$s, (b) 40~$\mu$s, and (c) 60~$\mu$s. The images were obtained by averaging ten measurements. The measured optical depth (OD) of the sample decreases as $\tau_0$ increases due to the population depletion of the $F=1$ spin states induced by the probe beam.}
\label{figure1}
\end{figure}

In Sec.~II, we provide a brief description of $F=1$ absorption imaging, and present the analytic solution of spin population evolution and the resultant  optical signal of absorption imaging. In Sec.~III, we verify the theoretical prediction by measuring the absorption spectra of $F=1$ $^{23}$Na BECs in various spin states. Finally, in Sec.~IV, we provide a summary of this work.

\section{Theory}\label{section2}

\subsection{F=1 imaging}

We consider an imaging situation where a cloud of alkali atoms with $I=3/2$ nuclear spin are prepared in the $F=1$ hyperfine ground state and a $\sigma^{-}$-polarized probe beam is irradiated to the atoms for $D_2$ optical transition [Fig.~2(a)]. After passing the sample, the electric field, $E$, of the probe beam is expressed as
\begin{equation}
E= E_0~e^{ik \int dz \chi(z) /2},
\end{equation}
where $E_0$ is the electric field in the absence of the sample, $k$ is the wave number of the probe beam, and $\chi(z)$ is the electric susceptibility of the sample, which is determined by the details of the interaction between the atoms and the probe light and depends on the atomic spin state and the polarization and intensity of the probe light~\cite{Hecht}. Here, the $z$-direction is the propagation direction of the probe beam.

In our study, we assume an optically thin sample with uniform atomic spin states in the beam propagation direction, and a weak probe beam in the linear regime where $\chi$ is independent of the beam intensity. Furthermore, we ignore spin coherence between the Zeeman states in the $F=1$ hyperfine level, which is the case in a realistic experimental condition where the imaging duration is a few tens of $\mu$s, much longer than the lifetime of optically excited states.

For atoms in the $F=1$ hyperfine ground state, the $\sigma^-$-polarized probe beam can drive the optical dipole transition from the $|F_g=1,m=-1,0,1\rangle$ state to the $|F_e, m-1\rangle$ states with $|m-1|\leq F_e \leq 2$. Here we use $F_g$ and $F_e$ to denote specifically the hyperfine spin numbers of the ground state and the optically excited state, respectively. Including all of the possible dipole transitions, the electric susceptibility $\chi$ of the sample is given by~\cite{Loudon}
\begin{equation}
i k \int dz \chi(z) /2 = -\frac{\tilde{n} \sigma_0}{2} \sum_{m=-1}^{1} \sum_{F_e=|m-1|}^2  \frac{R_{1,m}^{F_e,m-1}}{1-2i \delta_{F_e} /\Gamma} f_m,
\end{equation}
where $\tilde{n}(x,y)$ is the column atomic density along the $z$ direction, $\sigma_0=3 \lambda^2/ (2\pi)$ is the resonant cross section of the atom ($\lambda$ is the wavelength of the resonant light), $R_{F_g,m_g}^{F_e,m_e}$ is the relative transition strength for the $|F_g,m_g\rangle \rightarrow |F_e,m_e\rangle$ transition with respect to the $|F_g=2,m=-2\rangle \rightarrow |F_e=3,m=-3\rangle$ cyclic transition [Fig.~2(b)]~\cite{footnote1},  $f_m$ is the fractional spin population of the $|F_g=1, m_z=m\rangle$ state, $\delta_{j}$ is the frequency detuning of the probe beam with respect to the $|F_g=1\rangle \rightarrow |F_e=j\rangle$ transition ($j=0,1,2$) [Fig.~2(a)], and $\Gamma$ is the natural linewidth of the optically excited states. From Eqs.~(1) and (2), and the relation $E/E_0=|t| e^{i\phi}$, the transmittance $|t|$ and phase shift $\phi$ of the probe beam are explicitly given by
\begin{eqnarray}
\ln |t| = -\frac{\tilde{n} \sigma_0}{24} &\big[&6 L_2 f_{-1}+(3 L_2+ 5 L_1)f_0
\nonumber\\
&&+(L_2 + 5 L_1+ 4 L_0)f_1 \big],
\nonumber\\
\phi = - \frac{\tilde{n} \sigma_0}{24}&\big[& 6 D_2 f_{-1}+(3 D_2 +5 D_1)f_0
\nonumber\\
&&+(D_2+ 5 D_1+ 4 D_0)f_1 \big],
\label{eq4}
\end{eqnarray}
where $L_j = 1/[1+4(\delta_j/\Gamma)^2]$ and $D_j =2 (\delta_j / \Gamma)/[1+4(\delta_j/\Gamma)^2]$.

\begin{figure}
\centering
\includegraphics[width=7.6cm]{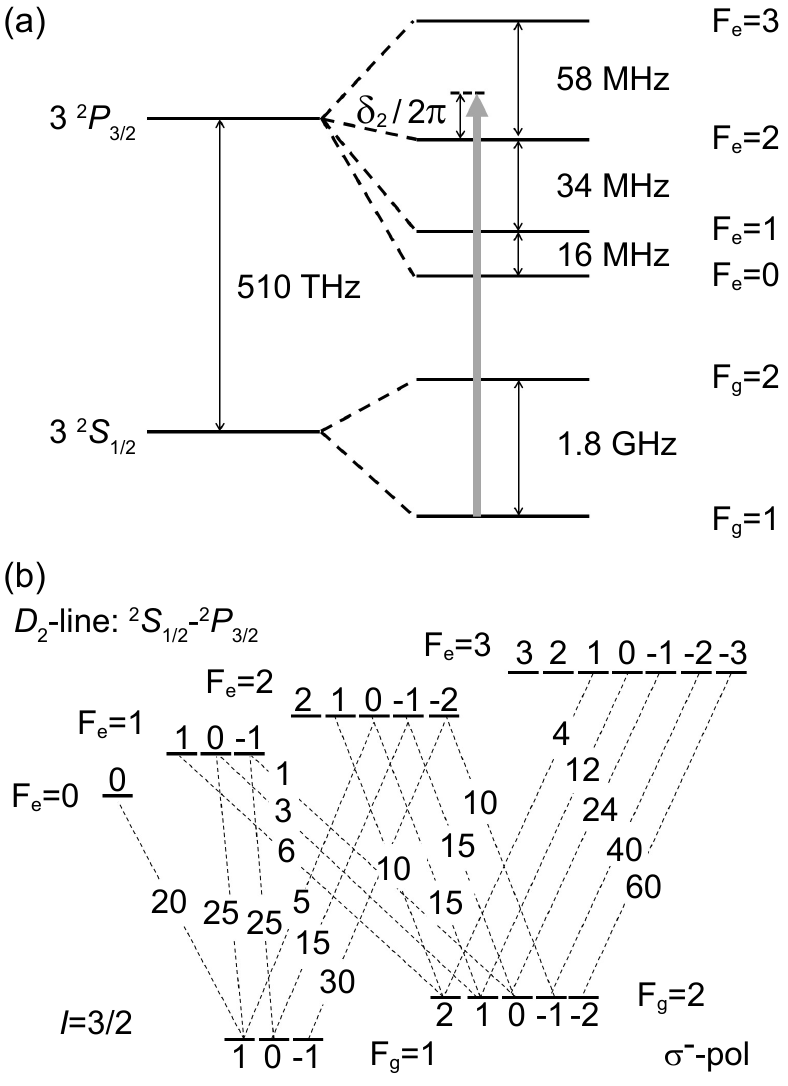}
\caption{(a) Energy levels of a $^{23}$Na atom. The gray arrow indicates the optical driving of the probe beam in $F=1$ imaging and $\delta_2$ is the frequency detuning of the probe beam with respect to the $|F_g=1\rangle\rightarrow |F_e=2\rangle$ transition. (b) Relative strengths of D$_2$ optical transitions of an atom with $I=3/2$ nuclear spin for $\sigma^{-}$-polarized light~\cite{Metcalf}. For convenience, the strength of the cyclic transition $|F_g=2, m=-2 \rangle \rightarrow |F_e=3, m=3\rangle$ is set to be 60.}
\label{figure2}
\end {figure}

In absorption imaging, the two-dimensional distribution of the optical depth (OD) of the sample is determined by $\mathrm{OD}(x,y)=-\ln |t|^2=-\ln(I/I_0)$, where the intensities $I$ and $I_0$ of the probe beam are measured with and without the sample, respectively. In an experiment, when a probe beam has a finite pulse duration $\tau_0$, the measured OD of the sample is interpreted as
\begin{eqnarray}
\mathrm{OD}&=&-\ln\Big[ \Big( \int^{\tau_0}_0 I(\tau) d\tau \Big) /\Big( \int^{\tau_0}_0 I_0(\tau) d\tau \Big) \Big] \\
&=&-\ln\Big[ \Big( \int^{\tau_0}_0 \big|t(\tau)\big|^2 I_0(\tau) d\tau \Big) /\Big( \int^{\tau_0}_0 I_0(\tau) d\tau \Big) \Big].
\label{OD}
\end{eqnarray}
If the fractional spin populations, $\{f_m\}$, of the $F_g=1$ hyperfine level change during imaging because of the optical pumping effect, the transmittance would temporally vary as $t(f_m(\tau))$. Thus, the measured OD is dependent on the probe-beam pulse duration in the absorption image. In the limit $\tau_0 \rightarrow 0$, the measured OD approaches the true sample OD in an initial condition.

\subsection{Optical pumping effect}

The evolution of the atomic spin state can be described by a set of rate equations for the spin populations~\cite{Meystre}. Including all of the contributions from the dipole transitions driven by the probe beam and spontaneous decays of the excited states, the rate equations can be written as
\begin{eqnarray}
\frac{d f_m}{d\tau} &=& -\sum_{F_e=0}^{2} R_{1,m}^{F_e,m-1} \frac{\Gamma}{2}s_0 L_{F_e} (f_m-f^{F_e}_{m-1})
\nonumber\\
&&+ \sum_{F_e=0}^{2} \sum_{m_e=m-1}^{m+1} \Gamma R_{1,m}^{F_e,m_e} f^{F_e}_{m_e},
\\
\frac{d f^{F_e}_m}{d\tau} &=& R_{1,m+1}^{F_e,m} \frac{\Gamma}{2}s_0 L_{F_e} (f_{m+1}-f^{F_e}_m )-\Gamma f^{F_e}_m,
\label{Moon2}
\end{eqnarray}
where $s_0=I_0/I_s$ is the on-resonance saturation parameter, $I_s=\pi h c \Gamma/(3\lambda^3)$ is the saturation intensity for optical transitions, $h$ is the Planck constant, $c$ is the speed of light, and $f^{F_e}_{m}$ is the fractional spin population of the  $|F_e,m\rangle$ state. Here, we assume that the $F_g=2$ hyperfine states are completely dark to the probe beam, and once the atoms decay into the $F_g=2$ state via optically excited states, they cannot return to the $F_g=1$ state.

In the weak probing beam limit $s_0\ll 1$, following the discussion described in Refs.~\cite{Moon2, Moon3}, we can obtain an analytic solution of the spin population evolution. Setting $d{f}^{F_e}_m/{d\tau}=0$ for $s_0\ll 1$, from Eq.~(7) we have $f^{F_e}_m = R_{1,m+1}^{F_e,m} \frac{s_0}{2}L_{F_e} f_{m+1}$ . Then, an adiabatic elimination of $f^{F_e}_{m_e}$ in Eq.~(6) gives 
\begin{eqnarray}
\frac{d {f}_m}{d\tau} &=& -\sum_{F_e=0}^{2} R_{1,m}^{F_e,m-1} \frac{\Gamma}{2}s_0 L_{F_e} f_m
\nonumber\\
&&+ \sum_{F_e=0}^{2} \sum_{m'=m-1}^{m+1} R_{1,m}^{F_e,m'} R_{1,m'+1}^{F_e,m'}\frac{\Gamma}{2}s_0 L_{F_e} f_{m'+1}.
\label{Rate}
\end{eqnarray}
In general, $s_0$ is proportional to the incident beam intensity and $I_0$ varies during imaging. Introducing a new quantity $\tilde{\tau}=\frac{\Gamma}{2}\int^\tau_0 s_0(\tau') d\tau'$, we can rewrite the rate equations in dimensionless form as
\begin{eqnarray}
\frac{d f_m}{d \tilde{\tau}} &=& -\sum_{F_e=0}^{2} R_{1,m}^{F_e,m-1}  L_{F_e} f_m
\nonumber\\
&+& \sum_{F_e=0}^{2} \sum_{m'=m-1}^{m+1} R_{1,m}^{F_e,m'} R_{1,m'+1}^{F_e,m'} L_{F_e} f_{m'+1}.
\end{eqnarray}
The quantity $\tilde{\tau}$ can be regarded as the accumulated magnitude of the driving by the probe beam.  

Then, we obtain the analytic solution of the spin populations $f_m(\tau)$ to Eq.~(9) as~\cite{Moon2, Moon3}
\begin{eqnarray}
f_{-1}(\tilde{\tau})&=& f_{-1}^0 e^{-\lambda_{-1} \tilde{\tau}} \nonumber \\
&+& C_{-1}^{0}(f_0^0 - C_{0}^{1}f_1^0)(e^{-\lambda_0 \tilde{\tau}}- e^{-\lambda_{-1}  \tilde{\tau}})
\nonumber\\
&+& C_{-1}^{1} f_1^0 (e^{-\lambda_1 \tilde{\tau}}- e^{-\lambda_{-1} \tilde{\tau}}),
\nonumber\\
f_0(\tilde{\tau})&=&f_0^0 e^{-\lambda_0 \tilde{\tau}} + C_{0}^{1} f_1^0 (e^{-\lambda_1  \tilde{\tau}}-e^{-\lambda_0  \tilde{\tau}}),
\nonumber\\
f_1(\tilde{\tau})&=& f_1^0 e^{-\lambda_1 \tilde{\tau}},
\label{Noh1}
\end{eqnarray}
where $f_m^0$ are the initial values of $f_m$ at $\tau=0$,
\begin{eqnarray}
\lambda_{-1}&=& \dfrac{1}{4}L_2,\nonumber\\
\lambda_{0} &=& \dfrac{3}{16}L_2+\dfrac{35}{144}L_1,\nonumber\\
\lambda_{1} &=& \dfrac{11}{144}L_2+\dfrac{35}{144}L_1+\dfrac{2}{9}L_0,
\label{Noh2}
\end{eqnarray}
and
\begin{eqnarray}
C_{-1}^{0}&=&\dfrac{9L_2+25L_1}{9L_2-35L_1},\nonumber\\
C_{-1}^{1}&=&\dfrac{13L_2^2+125L_2 L_1+92L_2 L_0-100L_1 L_0-128L_0^2}{4(L_2-2L_0)(25L_2-35L_1-32L_0)},\nonumber\\
C_{0}^{1}&=&\dfrac{L_2+4L_0}{4(L_2-2L_0)}.
\end{eqnarray}
$\lambda_m$ is the depletion rate of the $|F_g=1,m\rangle$ state and $C_m^{m'}$ represents the population transfer from the $|F_g=1,m'\rangle$ state to the $|F_g=1,m\rangle$ state.

\begin{figure}
\centering
\includegraphics[width=8.4cm]{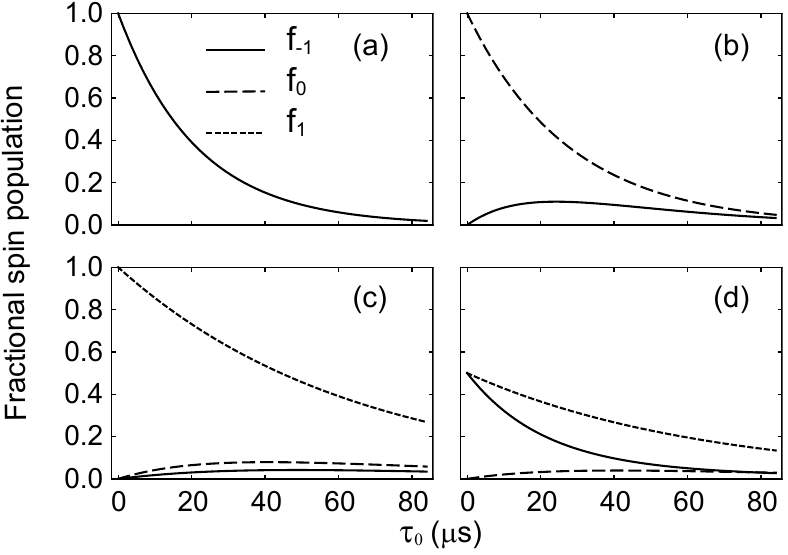}
\caption{Time evolutions of the spin populations under a $\sigma^{-}$-polarized probing light for various initial spin conditions: (a) $(f_{-1}^0,f_0^0,f_1^0)=(1,0,0)$, (b) $(0,1,0)$, (c) $(0,0,-1)$, and (d) $(1/2,0,1/2)$. The probe beam is set to be on resonance to the $|F_g=1\rangle \rightarrow |F_e=2\rangle$ transition, i.e., $\delta_2=0$; and the saturation parameter of the probe beam is $s_0=0.006$.}
\label{figure2}
\end {figure}

\begin{figure*}
	\centering
	\includegraphics[width=17cm]{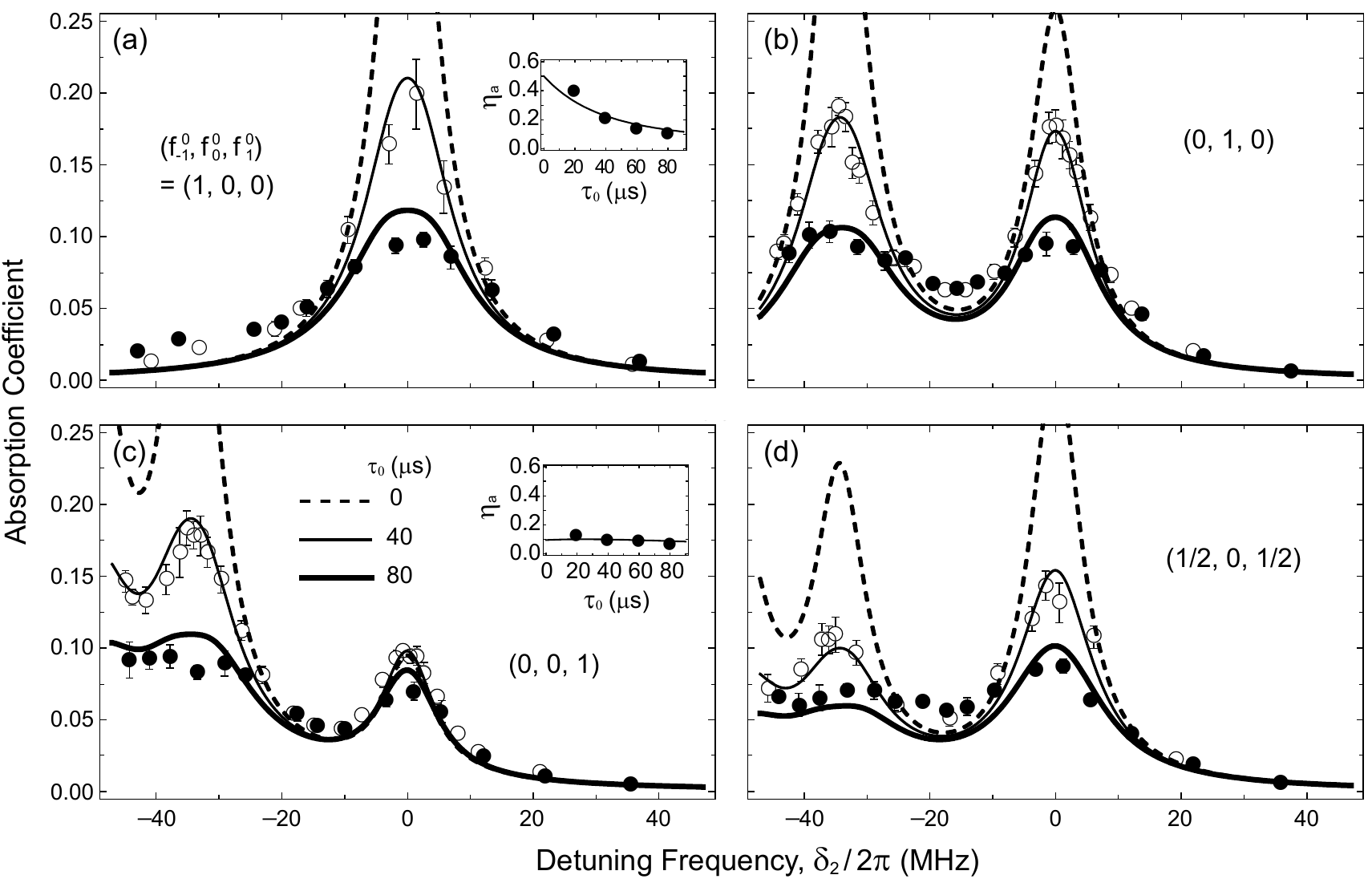}
	\caption{Absorption spectra of BECs of $^{23}$Na atoms in the $F_g=1$ hyperfine level. The initial spin states of the condensate were (a) $(f_{-1}^0,f_0^0,f_{1}^0)=(1,0,0)$, (b) $(0,1,0)$, (c) $(0,0,1)$, and (d) $(1/2,0,1/2)$. The saturation parameter of the probe beam was $s_0=6.1\times 10^{-3}$, and the probe beam pulse duration was $\tau_0=40~\mu$s (open circles) and 80~$\mu$s (solid circles). Each data point was obtained from ten measurements and its error bar indicates the standard deviation of the measurements. The thick,  and thin lines are the theoretical predictions of Eq.~(16) for $\tau_0=40~\mu$s and 20~$\mu$s, respectively, and the dashed lines indicate the spectra in the limit of $\tau_0\rightarrow 0~\mu$s. The insets in (a) and (c) show the absorption coefficient $\eta_a$ as a function of $\tau_0$, measured with $\delta_2=0$ for $(f_{-1}^0,f_0^0,f_{1}^0)=(1,0,0)$ and $(0,0,1)$, respectively. The lines are obtained from the analytic expression in Eq.~(16).}
	\label{figure3}
	\end {figure*}

In Fig. 3, we display several examples of spin population evolution under the probe light. It is clearly seen that the spin populations are transferred to lower $m$ states in $\sigma^{-}$-polarized light. When the sample is initially prepared in the $|m=1\rangle$ state, the populations of the $|m=-1\rangle$ and $|m=0\rangle$ states grow while that of the $|m=1\rangle$ state undergoes a simple exponential decay [Fig.~3(c)]. For an equal mixture of the $m=-1$ and $m=1$ spin components, we observe different decay rates for each spin component as shown in Fig.~3(d).

\subsection{Absorption imaging signal}

With the time-varying spin populations, $f_m(\tilde{\tau})$, we can numerically evaluate Eq.~(4) to determine the optical signal of absorption imaging. Using the relation $\tilde{\tau}=\Gamma/(2 I_s) \int^\tau_0 I_0(\tau') d\tau'$, we have
\begin{eqnarray}
\int^{\tau_0}_0 I(\tau) d\tau &=&\int^{\tau_0}_0 \big|t(\{f_m\})\big|^2 I_0(\tau) d\tau
\\ &=&\frac{2I_s}{\Gamma}\int^{\tilde{\tau}_0}_0 \big|t(\{f_m(\tilde{\tau})\})\big|^2 d\tilde{\tau},
\end{eqnarray}
and thus we obtain
\begin{equation}
\mathrm{OD}(\tilde{\tau}_0;\{f^0_m\})=-\ln\Big[\frac{1}{\tilde{\tau}_0}\int^{\tilde{\tau}_0}_0 \big|t(\{f_m(\tilde{\tau}))\})\big|^2 d\tilde{\tau}\Big].
\end{equation}
Note that the OD measured with a weak intensity probe beam is determined only by $\tilde{\tau}_0$ and independent of the probe beam pulse shape. This can be intuitively understood by noting that $\tilde{\tau}_0$ is proportional to the total number of photons irradiated on the sample and that the OD is directly related to the average probability for the photons to be absorbed by the sample. When the saturation effect is not negligible at high intensities, the OD of the sample becomes sensitive to the probe beam pulse shape~\cite{Kwon12}.

When the optical pumping effect is not significant, i.e., $\lambda_m\tilde{\tau}_0 \ll 1$, the imaging signal can be well estimated by the averaged spin populations during imaging, $\bar{f}_m=\frac{1}{\tilde{\tau}_0} \int^{\tilde{\tau}_0}_0 f_m(\tilde{\tau}) d\tilde{\tau}$. Then, the OD is explicitly expressed as
\begin{eqnarray}
\mathrm{OD} = \frac{\tilde{n} \sigma_0}{12} &\big[&6 L_2 \bar{f}_{-1}+(3 L_2+ 5 L_1)\bar{f}_0
\nonumber\\
&&+(L_2 + 5 L_1+ 4 L_0)\bar{f}_1 \big],
\end{eqnarray}
with
\begin{eqnarray}
\bar{f}_{-1} &=& f_{-1}^0 E(\lambda_{-1}\tilde{\tau}_0) \nonumber \\
&+&C_{-1}^{0}(f_0^0 - C_{0}^{1}f_1^0)[E(\lambda_0\tilde{\tau}_0)-E(\lambda_{-1}\tilde{\tau}_0)]
\nonumber\\
&+& C_{-1}^{1} f_1^0 [E(\lambda_{1}\tilde{\tau}_0)-E(\lambda_{-1}\tilde{\tau}_0)],
\nonumber\\
\bar{f}_0 &=& f_0^0 E(\lambda_0\tilde{\tau}_0)+ C_{0}^{1} f_1^0 [E(\lambda_{1}\tilde{\tau}_0)-E(\lambda_0\tilde{\tau}_0)],
\nonumber\\
\bar{f}_1 &=& f_1^0 E(\lambda_{1}\tilde{\tau}_0),
\label{Noh5}
\end{eqnarray}
where $E(x)\equiv(1-e^{-x})/x$. This analytic expression of the OD is the main result of this work.

\section{Experiments}


To verify the analytic result for the absorption imaging signal, we experimentally measured the absorption spectra of $F=1$ $^{23}$Na BECs in various spin states. A BEC of $^{23}$Na atoms in the $|F_g=1,m=-1\rangle$ hyperfine spin state was first generated in an optically plugged magnetic quadrupole trap~\cite{Heo} and then transferred to an optical dipole trap. After deep evaporative cooling, an almost pure BEC with no discernible thermal atoms was prepared in the optical trap. The spin state of the BEC could be transmuted into other $F=1$ spin states using an adiabatic Landau-Zener rf sweep~\cite{Choi2}. When preparing a mixture of the $m=1$ and $m=-1$ spin components, we applied a $\pi/2$ pulse of rf field to the BEC in the $|m=0\rangle$ state~\cite{Seo}. Then, an $F=1$ absorption image of the sample was taken along the $z$ direction after a 24 ms time of flight. The probe beam was $\sigma^{-}$ polarized and its intensity was $38~\mu$W/cm$^2$, corresponding to $s_0=6.1\times10^{-3}$. When taking the image, an external magnetic field of $\sim 0.5$~G was applied along the $z$ direction.

We determined the absorption coefficient, $\eta_a=\mathrm{OD}/(\sigma_0 \tilde{n})$, of the sample from the image,  where $\tilde{n}$ is the column density estimated from the measured Thomas-Fermi (TF) radii of the condensate, $(R_x,R_y)\approx(154,119)~\mu$m, and the trapping frequencies of the optical trap, $(\omega_x,\omega_y,\omega_z)/(2\pi)=(4.1,5.3,510)$~Hz. In TF mean-field description, the column density distribution is given by $\tilde{n}(x,y)=\tilde{n}_0 (1-\frac{x^2}{R_x^2}-\frac{y^2}{R_y^2})^{3/2}$ with $\tilde{n}_0=\frac{m^2}{6\pi \hbar^2 a} (\omega_x \omega_y R_x R_y )^{3/2}/\omega_z\approx 8.4$, where $a$ is the scattering length of atoms. In the $F=2$ imaging where the atoms were optically pumped into the $|F_g=2, m=-2\rangle$ state~\cite{footnote} and imaged using the $|F_g=2,m=-2\rangle\rightarrow |F_e=3,m=-3\rangle$ cyclic transition, we checked that the imaging signal was consistent with the theoretical expectation of $\eta_a=1$.

Figure 4 shows the measurement results of the absorption coefficient $\eta_a$ as functions of the frequency detuning $\delta_2$ of the probe beam from the $|F_g=1\rangle \rightarrow |F_e=2\rangle$ transition for various spin states and two different pulse durations, $\tau_0=40~\mu$s and 80~$\mu$s. We found that the results are in good quantitative agreement with the analytic expression in Eq.~(16). In the calculation, we ignored the Zeeman energy shift caused by the external magnetic field, which is less than a few MHz. The dashed lines indicate the expected spectra in the limit of $\tau_0\rightarrow 0$ where the optical pumping effect is absent. It is clearly shown that the spectral peak strength decreases with increasing pulse duration, especially around the resonances. We note that there are small but pronounced deviations of the measurement results from the prediction, in particular, on the negative detuning side in Fig.~4(a) and the regions around $\delta_2/2\pi=-20$~MHz in Figs.~4(b) and 4(d).  The insets of Figs.~3(a) and 3(c) show the absorption coefficient as a function of the pulse duration, $\tau_0$, for the samples in the $|m=-1\rangle$ and $|m=1\rangle$ states, respectively. They are also quantitatively well accounted for by Eq.~(16). In the case of the $m=1$ sample, $\eta_a$ appears to be insensitive to $\tau_0$ because of the population transfer to the $m=0$ and $m=-1$ states as shown in  Fig.~3(c).

\section{Summary}\label{section5}

We analytically and experimentally investigated the optical pumping effect in $F=1$ absorption imaging of $^{23}$Na atoms. We presented an analytic expression for the imaging signal as a function of the total integral of the incident beam intensity, and provided experimental verification of the result by measuring the absorption spectra for $F=1$ spinor BECs in various imaging conditions. 

Our analytic result is useful in the quantitative analysis of $F=1$ imaging. One immediate application is the correction of the systematic error caused by spatial inhomogeneity of the probe-beam intensity. If the probe-beam intensity is spatially inhomogeneous over the imaging area, the optical pumping effect would vary spatially, leading to spatial modulations of the OD in the absorption image. This systematic error can be compensated for by our formula of the OD, where the $\tilde{\tau}_0$ dependence is explicitly given. Another application is imaging of high-column-density samples with a high-$\tilde{\tau}_0$ probe beam. Because the OD decreases with increasing $\tilde{\tau}_0$, one can tune the OD range of the sample to the dynamic range of an imaging system. Moreover, we expect that the analytic solution of spin population evolution in the probing beam can be used in quantitative analysis of other $F=1$ imaging methods such as near-resonant phase-contrast imaging, which was employed in recent spinor BEC experiments to measure the magnetization distribution of the BEC~\cite{Seo}. In contrast to conventional phase-contrast imaging, photon absorption and the subsequent optical pumping effect would not be negligible in imaging with a near-resonant probe beam. Finally, we emphasize that our result can be applied to other alkali atoms with $I=3/2$, and readily used for image analysis in experiments with $^{87}$Rb atoms.

\begin{acknowledgments}
This work was supported by IBS-R009-D1 and the National Research Foundation of Korea (Grants No. 2013-H1A8A1003984 and No. 2014-R1A2A2A01006654).
\end{acknowledgments}


\begin{thebibliography}{1}
\bibitem{footnote1} $R_{F_g,m_g}^{F_e,m_e}= (2L_e+1)(2J_e+1)(2J_g+1)(2F_e+1)(2F_g+1)
\times \begin{bmatrix} \begin{Bmatrix} L_e & J_e & S \\ J_g & L_g & 1 \end{Bmatrix} \begin{Bmatrix} J_e & F_e & I \\ F_g & J_g & 1 \end{Bmatrix} \begin{pmatrix} F_g & 1 & F_e \\ m_g & m_e-m_g & -m_e\end{pmatrix}\end{bmatrix}^2$,
where $L$, $S$, $I$ are the quantum numbers of the orbital, electron spin, and nuclear spin angular momenta of the atom, respectively, and $\lbrace\rbrace$ and () denote the $6J$, $3J$ symbols.
\bibitem{footnote} During the $F=2$ imaging, a pumping light resonant on the $|F_g=1\rangle \rightarrow |F_e=2\rangle$ transition was irradiated to the condensate along the $x$ direction.
\end{thebibliography}
\end{document}